%% file: new_draft.tex
\documentclass[conference]{IEEEtran}
\IEEEoverridecommandlockouts
\usepackage{cite}
\usepackage{amsmath,amssymb,amsfonts}
\usepackage{algorithmic}
\usepackage{graphicx}
\usepackage{textcomp}
\usepackage{xcolor}
\usepackage[left=1.62cm,right=1.62cm,top=1.9cm]{geometry}
\usepackage{amsmath,amssymb,amsfonts,amsthm}
\usepackage{graphicx}
\usepackage{bbm}
\usepackage{xcolor}
\usepackage{circuitikz}
\usetikzlibrary{matrix, fit, positioning, calc, shapes.arrows}
\usepackage{tikz}
\usetikzlibrary{positioning}
\usetikzlibrary{shapes.gates.logic.US}
\newcounter{thm}

\newtheorem{definition}[thm]{Definition}

\newcommand*{\rom}[1]{\uppercase\expandafter{\romannumeral #1\relax}}

\newcommand{\ket}[1]{{|#1\rangle}}

\allowdisplaybreaks

\makeatletter

\counterwithout{equation}{section} % Remove the resetting by `section`
\def\BibTeX{{\rm B\kern-.05em{\sc i\kern-.025em b}\kern-.08em
    T\kern-.1667em\lower.7ex\hbox{E}\kern-.125emX}}
\begin{document}

\title{Comparing Latency and Power Consumption: Quantum vs. Classical Preprocessing

}

\author{
   \IEEEauthorblockN{Zuhra Amiri\IEEEauthorrefmark{1}, Janis N\"otzel\IEEEauthorrefmark{1}}
   \IEEEauthorblockA{
       \IEEEauthorrefmark{1}Emmy-Noether Group Theoretical Quantum Systems Design Lehrstuhl f\"ur Theoretische Informationstechnik,\\ Technische Universit\"at M\"unchen\\ 
  \{zuhra.amiri, janis.noetzel\}@tum.de 
    }}

\maketitle

\begin{abstract}
  Low latency and low power consumption are the main goals for our future networks. Fiber optics are already widely used for their faster speed. We want to investigate if optical decoding has further advantages to reaching future goals. We have investigated and compared the decoding latency and power consumption of an optical chip and its electronic counterpart built with MOSFETs. We have found that optical processing has a speed and power consumption benefit. For future networks and real-time applications, this can bring huge advantages over our current electronic processors.
\end{abstract}


\section{Introduction}

In optical computing, researchers have been trying to find ways to process information, with light being the primary resource. Interest was first sparked in the 1960s with the discovery of lasers. In the 1980s, essential logic functions could be performed with optical devices. Nowadays, advances in photonics have enabled the development of photonic integrated circuits (PICs), e.g., by Quix Quantum \cite{Quix}. These integrated circuits combine optical components and electronics on a single chip, allowing for efficient optical signal processing and computation. Optical processing can also benefit optical communication, especially for high-speed communication with high baud rates. One prominent application of this technology is in the Internet of Things (IoT) field, where optical processing can significantly enhance various aspects of data transmission and computation \cite{opticalIoT}.

One application where optical processing can be beneficial is the computation of the Walsh-Hadamard transform (WHT). The WHT is a robust algorithm widely employed in various domains, including data compression, image and audio processing, and error detection and correction. Its applications in these fields are crucial for ensuring efficient data representation, analysis, and transmission. Error detection and correction, in particular, play a vital role in digital communication systems where reliable data transmission is essential. Employing the error detection techniques based on the Walsh-Hadamard-Transform assure the integrity and accuracy of transmitted data, mitigating the adverse effects of noise and interference that can introduce errors.

One notable variant of the Walsh-Hadamard transform is the Fast Walsh-Hadamard Transform (FWHT). The FWHT is a computationally efficient version of the WHT that retains the advantages of the original transform while significantly reducing computational complexity. The FWHT has been widely recognized for its efficacy and has found practical applications in various fields. The potential of optical methods for implementing the FWHT has been explored by Guha, as demonstrated in his research paper \cite{Guha_2011}. Optical computation of the FWHT leverages the principles of butterfly operations and can be achieved using components such as beamsplitters. One of the notable advantages of optical implementation is its ability to asymptotically approach the Holevo capacity, which measures the maximum amount of information that can be reliably transmitted through a communication channel. The Holevo capacity can be achieved due to the superadditivity of the joint detection receiver \cite{Guha_2011, rosati2016}. Superadditivity can be achieved in quantum channels and is not achievable in classical communications \cite{Koudia_2022}.

An illustrative example of the electronic employment of the Hadamard code, based on the FWHT, is NASA's 1971 Mariner 9 mission \cite{NASA}. The Hadamard code was instrumental in correcting picture transmission errors, enabling the reception of high-resolution images of the Martian surface. This groundbreaking contribution revolutionized our understanding of Mars and highlighted the significance of error correction techniques in space exploration. 

 Our study aims to investigate whether an optical joint detection receiver offers distinct advantages over its electronic counterpart regarding latency and power consumption. These factors are paramount in modern IoT networks, where low latency enables real-time interactions, enhances user experience, and supports time-sensitive applications. Moreover, low power consumption is a crucial goal for future IoT networks to ensure sustainability and enable connectivity in remote areas with limited power sources \cite{opticalIoT}.
 
Low power consumption and low latency are also crucial for future networks \cite{6Ggoals}, as it saves costs and allows mobile and remote deployment to enable connectivity in rural and underdeveloped areas with limited power sources. This aspect is also crucial in deep-space communication, as only limited power is available. Low power consumption will be essential for future networks to ensure sustainability, reducing costs, and scalability.

To facilitate this comparison, we propose the design of a circuit that utilizes MOSFETs (Metal-Oxide-Semiconductor Field-Effect Transistors) to implement the FWHT. MOSFETs are widely regarded as an essential building block in modern electronics, owing to their desirable characteristics such as high speed, efficiency, and low power consumption. Our circuit design aims to harness the benefits of MOSFET technology in implementing the joint detection receiver. 

\section{Methodology}

The Hadamard code is based on the symmetric matrix $H_n$ and is defined as
    \begin{equation}
        (H_n)_{j,k} = (-1)^{j\cdot k},\quad j \cdot k = \textstyle\sum_{t=0}^{\log_2 n} j_t k_t,
    \end{equation}
    with $j \cdot k$ being the bitwise scalar product of the binary representations of $j,k = 0,...,n-1$ \cite{rosati2016}. Each row of the matrix represents a codeword. 
    Optically, the sender encodes their message into coherent states $|\alpha\rangle=\exp(-|\alpha|^2/2)\sum_{n=0}^\infty\alpha^n/\sqrt{n!}|n\rangle$, where $\{\ket{n}\}_{n\in\mathbb N}$ is the photon number basis of $\mathcal F$. For binary phase shift keying (BPSK), the set of signals states is $S_1=\{\ket{\alpha},\ket{-\alpha}\}$.
    
    The receiver performs the FWHT to decode the message. The FWHT relies on butterfly operations, which can be implemented optically with beamsplitters. 
Quantum-mechanically, the beamsplitter will transmit or reflect each incoming photon with some probability. It can be defined as follows:

    \begin{definition}[Beamsplitter]
A lossless two-mode beamsplitter, with two input and two output ports, in quantum optics is described by the unitary matrix $U_{BS}$, which has the form
\begin{align}
    \begin{pmatrix}
    \hat{b}_{1}\\
 \hat{b}_{2}
    \end{pmatrix} = U_{BS} \begin{pmatrix}
     \hat{a}_{1}\\
 \hat{a}_{2}
    \end{pmatrix}, \quad U_{BS} =  \begin{pmatrix}
     \sqrt{T} & e^{i\phi}\sqrt{R}\\
 -e^{-i\phi}\sqrt{R} & \sqrt{T}
    \end{pmatrix},
\end{align}
where the two input ports are represented by the 1st and 2nd mode annihilation operators, $\hat{a}_{1},\hat{a}_{2}$, respectively, and the two output ports by $\hat{b}_{1}, \hat{b}_{2}$.  Furthermore $T$ and $R$ are the transmittance and  reflectance with $R+T = 1$, and $\phi$ is the phaseshift \cite{makarov2022theory}. 
\end{definition}
    In our work $50:50$ beamsplitters are used, which means that $R = T = 1/ \sqrt{2}$ and the phaseshift is $\phi = \pi/2$.
    
    We initially converted the optical model into a digital representation to compare the optical model with the electronic model. To accomplish this, we designated $\ket{\alpha}$ to be represented by the binary code $01$, $\ket{-\alpha}$ by $10$, and the vacuum state as $00$. For simplicity, we disregarded the signal amplification by the beamsplitter in the digital model. Table \ref{tab:binary bs} 
presents the truth table of the electronic model of the beamsplitter. 
    
    \begin{table}[]
    \centering
    \begin{tabular}{|c|c|}
   \hline
        $A_1$ $A_2$, $B_1$ $B_2$ & $C_1$ $C_2$, $D_1$ $D_2$\\
        \hline
         00\ 00\ & 00\ 00\\
        \hline
        01\ 01\ &\ 01\ 00\\
        01\ 10\ &\ 00\ 01\\
        \hline
        10\ 01\ &\ 00\ 10\\
        10\ 10\ &\ 10\ 00\\
       \hline
    \end{tabular}
    \caption{Binary beamsplitter transformations}
    \label{tab:binary bs}
\end{table}

\begin{figure}[!ht]
    \centering
    \input{logic_bs}
    \caption{The logical "beamsplitter" consisting of 4 AND gates}
    \label{fig:logical_beamsplitter}
\end{figure}
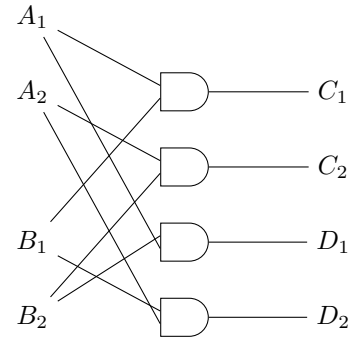

This truth table is realizable with 4 AND gates, as shown in Figure \ref{fig:logical_beamsplitter}.

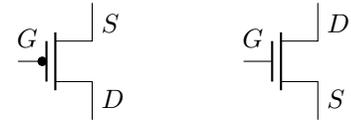
\begin{figure}[h!]
    \centering
    \input{pmosnmos}
    \caption{PMOS and NMOS structure}
    \label{fig: pmosnmos}
\end{figure}

Implementing logical gates with transistors is easily achievable in several ways. We chose the Metal-Oxide-Semiconductor Field-Effect Transistors (MOSFETs) that are the basis of the Complementary Metal-Oxide-Semiconductor (CMOS) technology since these transistors have low power consumption, high noise immunity, and high switching speeds \cite{Tietze}. The two CMOS transistors used to construct logic gates are the N-channel MOS (NMOS) and P-channel MOS (PMOS) transistors. In Figure \ref{fig: pmosnmos}, the structure of the MOSFETs is depicted. $G$ stands for gate, $S$ stands for source, and $D$ stands for drain. 

PMOS transistors allow current to flow between the source and drain terminals when a negative voltage is applied to the gate terminal. Conversely, NMOS transistors enable current to flow between the source and drain terminals when a positive voltage is applied to the gate terminal. 
To build the AND gate that we need, we have to use a NAND gate and a NOT gate because directly designing an AND gate with MOSFETs could result in problems, such as not getting high enough outputs, causing the following transistors to malfunction \cite{Andreou14}.

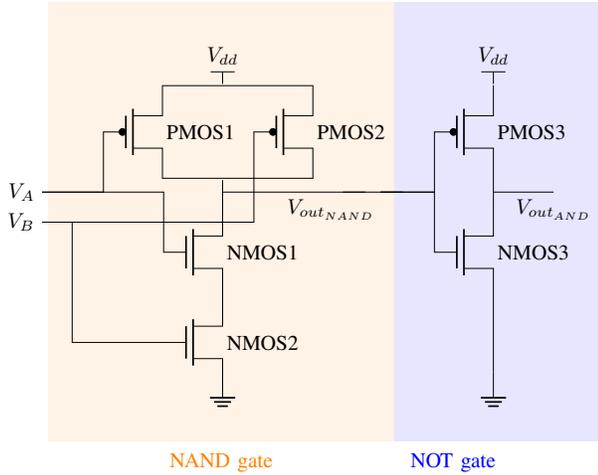
\begin{figure}[!ht]
    \centering
     \scalebox{.8}{\input{and-gate}}
    \caption{The AND-gate built with CMOS transistors}
    \label{fig:and with mosfet}
\end{figure}

Figure \ref{fig:and with mosfet} depicts the AND gate implemented with CMOS transistors. $V_A$ and $V_B$ are the inputs, and the output of the circuit is $V_{{out_{AND}}}$, which is the result of the AND gate. For computing the latency of the AND gate, it is essential to determine which inputs the NAND gate receives, as different transistors will be activated depending on this. 

If the inputs $V_A$ and $V_B$ are low, both PMOS transistors will switch on, while the NMOS transistor will remain switched off. 
The output of this circuit will be high, or $V_{dd}$, which will be the input of the NOT gate. Here the NMOS gate will switch on while PMOS remains off.   

If one of the inputs is high and the other low, then one of the PMOS and one of the NMOS switch on. Since only one NMOS is on, the ground is not connected. Hence the outcome will be high, which will be the input of the NOT gate. The NMOS switches on while PMOS stays off. 

If both inputs are high, both NMOS switch on, which means that the ground connects, giving us a low output. Therefore, the PMOS switches on in the NOT gate, giving us a high result.

Our goal is to compare the electronic version of the FWHT with the optical one regarding power consumption and decoding latency. For this, we will first introduce the methods of how we can calculate these. 

\subsection{Power Consumption}

It is important to take the power consumption into account since it's one of the main goals for future networks to reduce costs and be more energy efficient. These points are important for the environment, sustainability and scalability \cite{6Ggoals}. To compute the power consumption of the electronic FWHT, we have to take a look into the power consumption of the MOSFETs.  

A MOSFET operates in cutoff, saturation, and triode modes. For each of these, the power consumption differs depending on the gate-source voltage $V_{GS}$, the threshold voltage $V_{th}$ and the drain-source voltage $V_{DS}$ \cite[p. 178]{Tietze}.
$V_{GS}$ determines if the MOSFET is turned on or off. If the gate-source voltage is below the threshold voltage, the MOSFET stays off and the drain-source current $I_{D}$ doesn't flow. This means, the MOSFET is in cutoff mode ($V_{GS} < V_{th}$), in which the power consumption of the MOSFET is zero since the voltage applied to the gate is insufficient to turn it on. %\jn{what is $V_{GS}$ and what is $V_{th}$ ?}{} 

Only when the gate-source voltage is above the threshold voltage the MOSFET turns on. Now the drain-source voltage is important to determine in which mode the MOSFET operates. If 
$V_{GS} > V_{th}$ and $V_{DS} < V_{GS} - V_{th}$, meaning that if $V_{DS}$ is small, the MOSFET operates in triode mode. The computation of the power consumption in the triode region  is the least trivial. Firstly, we will need the definition of the drain current $I_d$ in triode mode
\begin{align}
    I_d = k [ (V_{GS}-V_{th}) V_{DS}-\tfrac{1}{2}V^2_{DS}], \label{eq: Idtriode}
\end{align}

with $k = \mu_n C_{ox}\frac{w}{l}$ and $V_{DS sat} =V_{GS} - V_{th}$ \cite{Wang_21}. In this region the drain current is dependent on both the gate-source voltage and the drain-source voltage. 

Furthermore, the datasheet of MOFETs gives information about the power dissipation for a fixed gate voltage $V_{GS}$. To calculate it for other voltages, one must consider the following definition, which can be derived from Eq. \eqref{eq: Idtriode} :
\begin{align}
   W =  \tfrac{k}{2}\left((V_{GS}-V_{th})(V^2_{DS}-V^2_{off})-\tfrac{1}{3}(V^3_{DS}-V^3_{off})\right). \label{eq:pd}
\end{align}

Under saturation mode ($V_{GS} > V_{th}$ and $V_{DS} \geq (V_{GS}-V_{th})$), the voltage applied to the gate is sufficient to turn the MOSFET entirely on. With the power formula $P = V \cdot I$, computation of the power consumption in this region is easy. The MOSFET datasheets contain the I-V curves, from which the power consumption can be derived. If this is not the case, the drain current in saturation mode is modeled as
\begin{align}
    I_d = \tfrac{k}{2} (V_{GS}-V_{th})^2 (1+ \lambda V_{DS}).
\end{align}

In this region the drain current is mostly independet on $V_{DS}$ as $\lambda$ is usually quite small. Unfortunately, most datasheets do not contain information about $k$ and $\lambda$. We can calculate $k$ with the given power dissipation from the datasheet to derive the power consumption in triode mode for other voltages than those shown in the datasheet. In \cite[p.180]{Tietze}, approximated coefficients for different MOSFETs can be found. 

%The power consumption and its loss should be low. \jn{Dieser Satz sollte besser in den Kontext eingebettet werden}{ist jetzt am anfang der section}

Since the beamsplitter is a passive optical device, no power will be consumed in the optical chip computation. However, since at the chip's input ports, the light's intensity should be the same, we need tunable beamsplitters before the actual computation of the Hadamard transform can take place. These devices will consume power for initial calibration.

 \subsection{Decoding Latency}
 Another important goal for future networks are low latency and high speed, reliable communication \cite{6Ggoals}. This is important for real-time applications, e.g. video conferencing, financial transactions and internet of things (IoT). 

Now that we have the devices for implementing the joint detection receiver both optically and electronically, we can calculate the latency. For the FWHT, we need $\frac{2^n}{2}\log 2^n$ beamsplitters.

For determining the switching speed of a MOSFET, it is crucial to consider the time required for the MOSFET to change from a conducting to a non-conducting state (or vice versa). This time depends on the time needed to charge or discharge the input and output capacitances of the MOSFET, which include the input capacitance ($C_{iss}$), output capacitance ($C_{oss}$), and reverse transfer capacitance ($C_{rss}$) \cite{Infineon}. %MOSFET datasheets contain this information.

%The capacitances are defined as follows:
%\begin{align}
 % C_{iss} &= C_{GS} + C_{DS} \quad \text{(with $C_{DS}$ shorted)}\\
  %  C_{rss} &= C_{GD}\\
 %C_{oss} &= C_{DS} +C_{GD},
%\end{align}

%with $C_{GS}$ being the gate-to-source capacitance, $C_{DS}$ being the drain-to-source capacitance, and $C_{GD}$ being the gate-to-drain capacitance \cite{Infineon}.
%The gate-to-drain capacitance ($C_{GD}$) is the most crucial parameter since it provides a feedback loop between the output and input of the circuit. It is also known as the Miller capacitance and is non-linear. 
It is important to note that the actual switching time of a MOSFET will depend on the specific gate charge of the MOSFET and that a higher gate charge will result in slower switching times.
Using different voltages, one can "choose" how fast a MOSFET can switch. It should take a manageable amount of time because that heats the device, which causes less efficiency. To calculate the switch times, we need the following definitions.

\begin{definition}
The turn-on delay $t_{d(on)}$ of a MOSFET is %given as
\begin{align}
    t_{d(on)}& = R_G C_{iss \text{ at } V_{DS}} \ln\left(\tfrac{V_{GS}}{V_{GS}-V_{gp}}\right) ,
   %R_G C_{iss \text{ at } V_{DS}}\left( \ln\left(\frac{1}{1-\frac{V_{TH}}{V_{GS}}}\right) +  \ln\left(\frac{V_{GS}-V_{TH}}{V_{GS}-V_{gp}}\right)\right),
\end{align}
with $R_G$ being the effective total gate resistance defined as the sum of internal gate resistance $R_g$ of the MOSFET and any external resistance $R_{gext}$ that is part of the gate drive circuitry. $V_{GS}$ is the voltage at the gate, $C_{iss}$ is the effective input capacitance of the MOSFET $C_{iss} = C_{gs} + C_{gd}$ and $V_{gp}$ is the gate plateau voltage \cite{Vishay_2016}.
\end{definition}

\begin{definition}
    The turn-off delay $t_{d(off)}$ is %given as
\begin{align}
    t_{d(off)} = R_G C_{iss \text{ at 0V}} \ln\left(\tfrac{V_{GS}}{V_{gp}}\right)
\end{align}
with $R_G$ being the effective total gate resistance of the MOSFET, $V_{GS}$ beging the voltage at the gate, $C_{iss}$ being the effective input capacitance of the MOSFET and $V_{gp}$ being the gate plateau voltage \cite{Vishay_2016}.
\end{definition}

It is noteworthy to note that datasheet switch times are often without external gate resistance. 

To calculate the decoding time for the optical counterpart, we need the size of the beamsplitter. The photons will pass through the beamsplitter with the speed of light.

\section{Results}
We have outlined our methodology for investigating the comparison of electrical and optical computation of the FWHT. Building upon the established methodology, we now turn our attention to the presentation of results and findings. In the following sections, we will delve into the results obtained from our analysis and explore their implications concerning our research questions and objectives.

Firstly, due to variations in transistor latency, we needed to select a specific transistor for our analysis. Among the MOSFET options available, we investigated the popular NMOS, SiRA04DP \cite{Vishay_NMOS}, and the popular PMOS, SiA469DJ \cite{Vishay_PMOS}. %As mentioned, t
The MOSFETs datasheets \cite{Vishay_NMOS, Vishay_PMOS} contain all the needed parameters to calculate the switch times and power consumption.

\subsection{Power Consumption}

As mentioned, power consumption differs in the three modes a MOSFET operates. 
In cutoff mode, the power consumption is $P = 0W$; in this mode, the MOSFET does not have enough power to operate. 

Then the MOSFET enters triode mode. To determine the power consumption in this mode, we must first calculate the drain current $I_d$. Then, we can determine the power consumption with the power formula $P = I_d * V_{DS}$ \cite[p.190]{Tietze}. In this phase, the MOSFET is turning on - the drain current rises as the drain-source voltage rises. 

\begin{figure}[!ht]
    \centering
    \includegraphics[scale = 0.17]{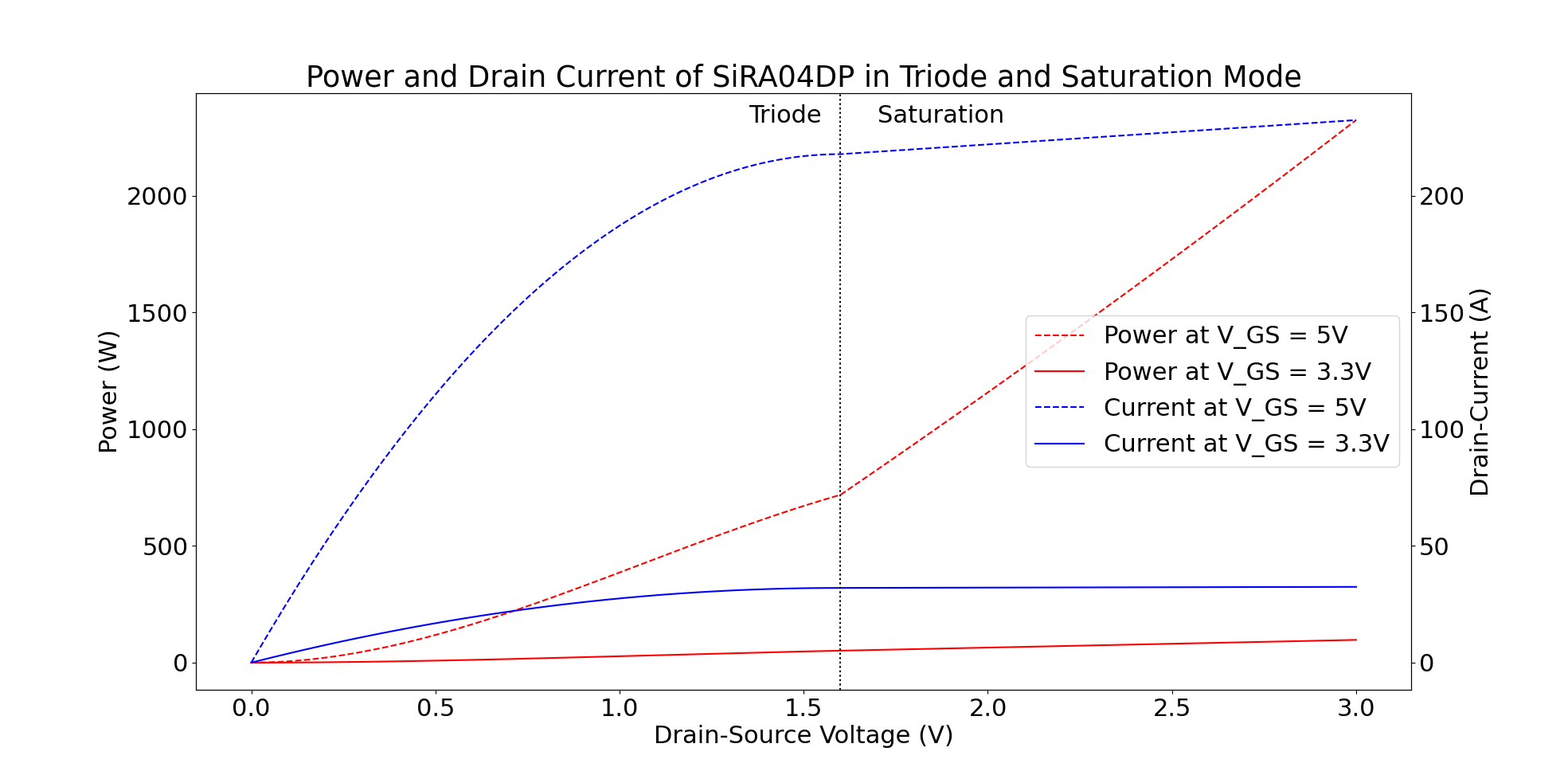}
    \caption{Power consumption of the NMOS in triode mode and saturation mode for $V_{GS} = 5V$ and $V_{GS} = 3.3V$}
    \label{fig:nmospowertriode}
\end{figure}

\begin{figure}[!ht]
    \centering
    \includegraphics[scale = 0.17]{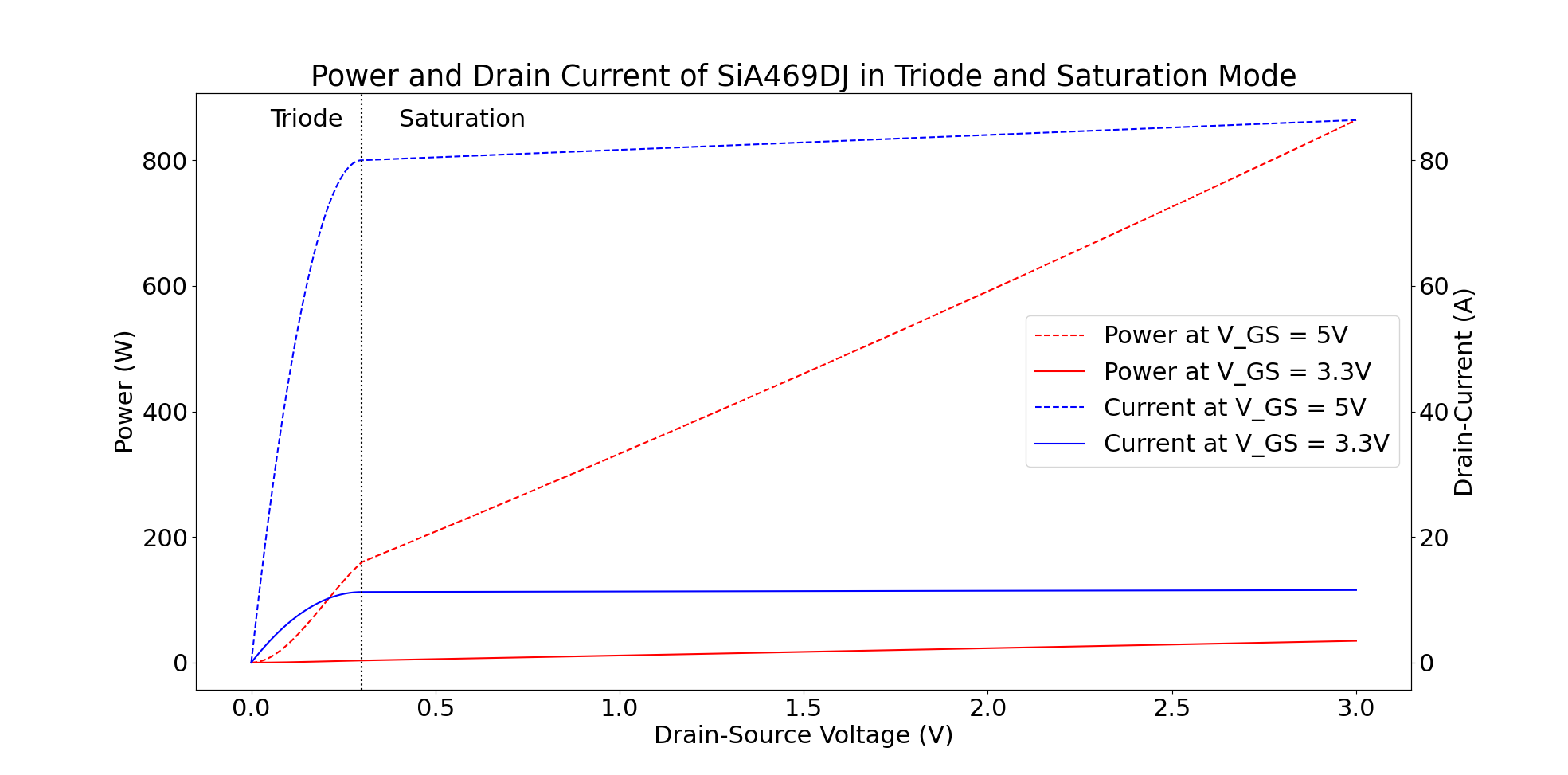}
    \caption{Power consumption of the PMOS in triode mode and saturation mode for $V_{GS} = 5V$ and $V_{GS} = 3.3V$}
    \label{fig:pmospowertriode}
\end{figure}

Lastly, the MOSFETs enter saturation mode, in which we need the same approach as above. 
Figures \ref{fig:nmospowertriode} and \ref{fig:pmospowertriode} depict the power consumption in triode mode and saturation mode for different drain-source voltages at $V_{GS} = 5V$ and $V_{GS} = 3.3V$ for the chosen NMOS and PMOS transistors. As one can see in saturation mode, the drain current becomes linear. 

It is visible that at $V_{GS} = 3.3V$, the power consumption of the MOSFET is significantly lower than at $5V$. The drain-current rises up to $35A$ for the NMOS at $V_{GS} = 3.3V$ and $235A$ at $V_{GS} = 5V$. For the PMOS, the drain current is lower. It rises up to $15A$ at $V_{GS} = 3.3V$ and $90A$ at $V_{GS} = 5V$.

In contrast, the beamsplitter is a passive device that does not need power, so its power consumption is $P = 0W$. The photonic chip by Quixquantum \cite{Quix} has tunable beamsplitters, which use power. However, this is only for the initial step to tune the beamsplitters. Afterward, they are passive and do not consume power. 

\subsection{Decoding Latency}

\begin{figure}[!ht]
    \centering
    \includegraphics[scale=0.21]{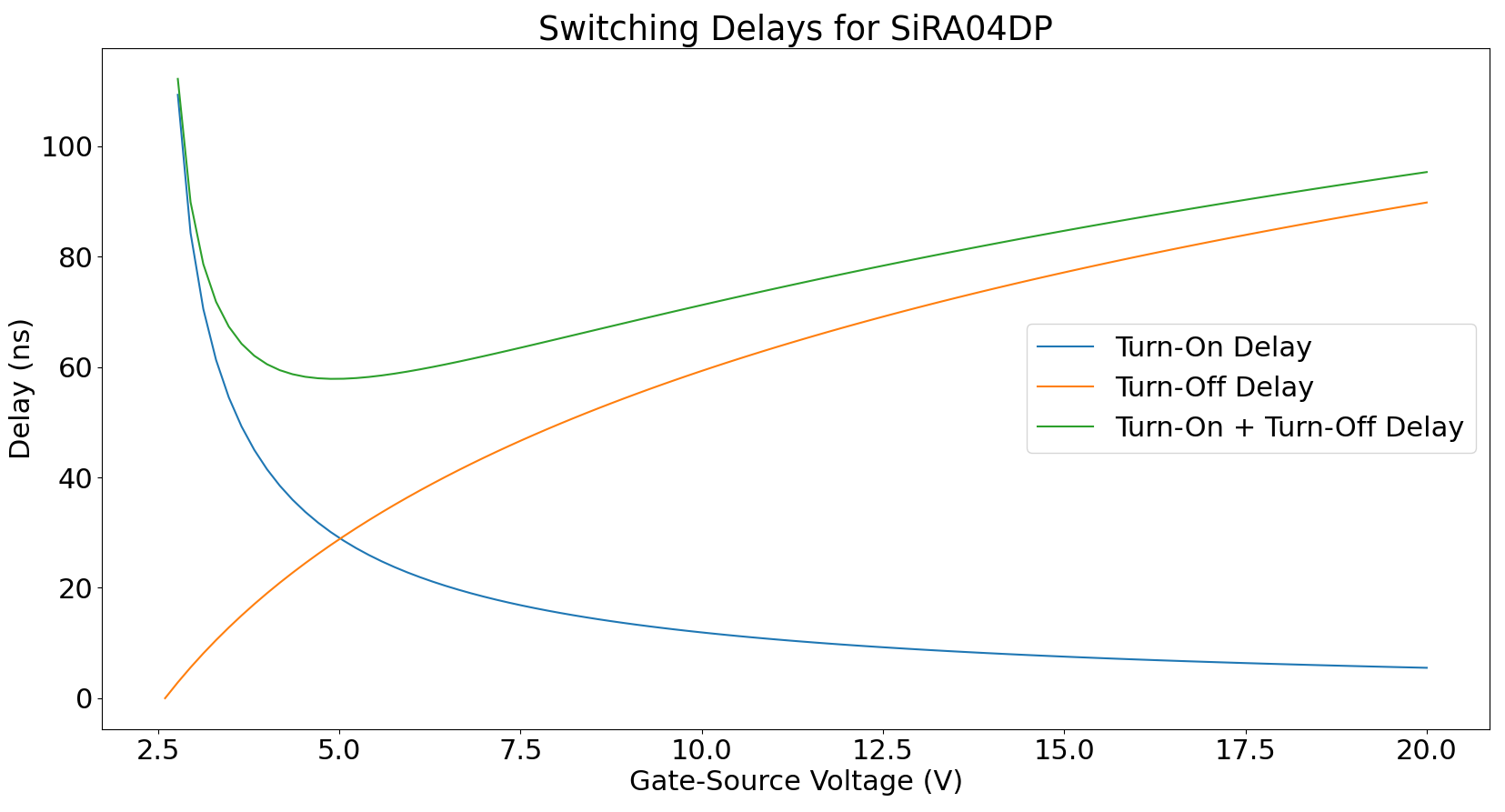}
    \caption{The turn on and turn off delay times of the NMOS SiRA04DP, with parameters $R_g = 1\Omega$, $C_{iss \text{ at} 0V} = 4000pF$, $C_{iss \text{at} V_{DS}} = 3600pF$, $V_{th} = 1.7 V$ and $V_{gp} =2.6V $ }
    \label{fig:nmosdelay}
\end{figure}

\begin{figure}[!ht]
    \centering
    \includegraphics[scale=0.21]{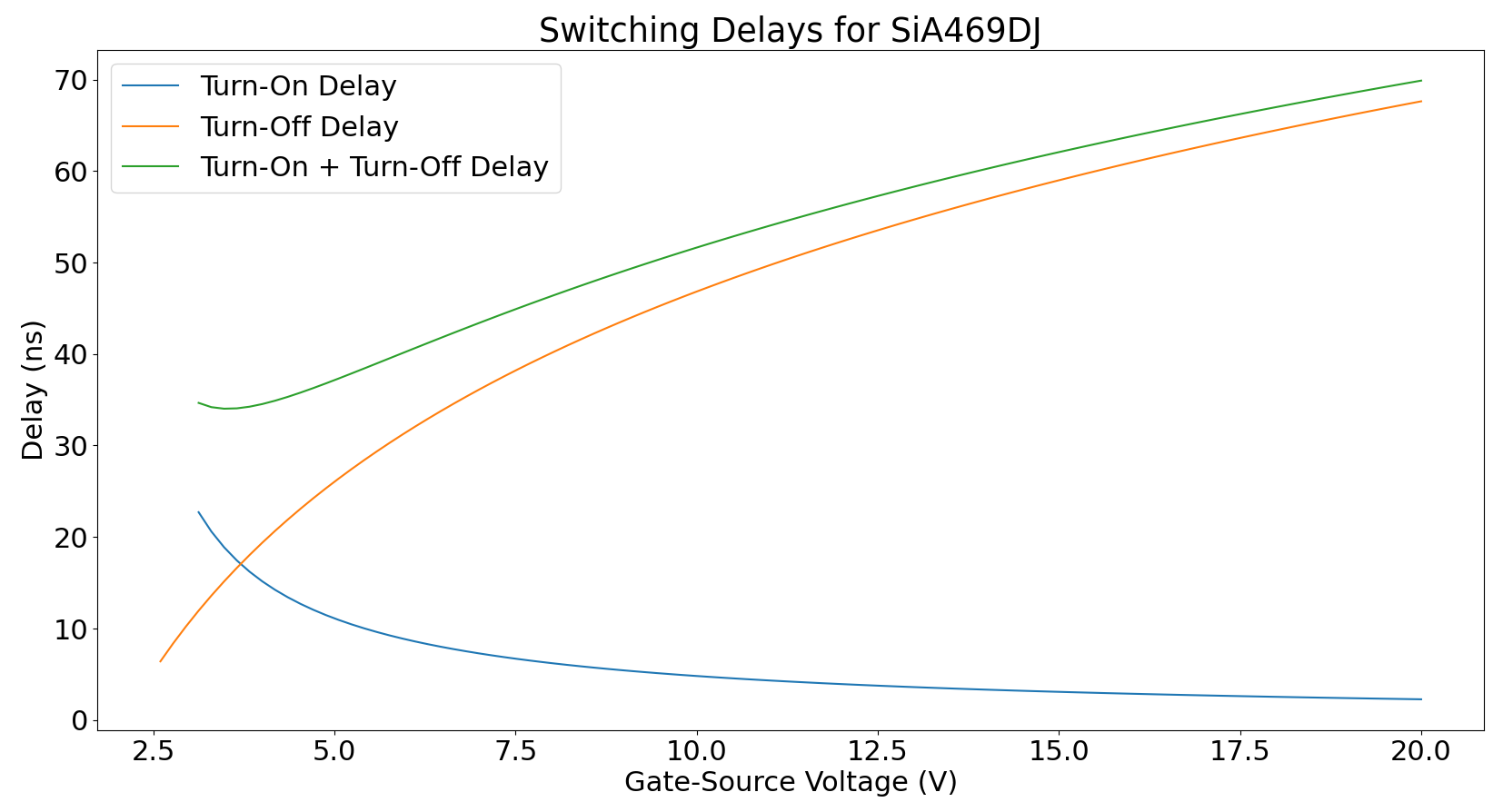}
    \caption{The turn on and turn off delay times of the PMOS SiA469DJ, with parameters $R_g = 9\Omega$, $C_{iss \text{ at} 0V} = 1500pF$, $C_{iss \text{at} V_{DS}}=1020pF$, $V_{th}=3V$ and $V_{gp}=2.1V$ }
    \label{fig:pmosdelay}
\end{figure}

To determine the decoding latency of the chosen MOSFETs, as mentioned before, an external gate resistor is crucial for the circuit design \cite{TexasInstruments}. Since we want fast switch times, we added a small external gate resistor of $10\Omega$. In Figure \ref{fig:nmosdelay} and Figure \ref{fig:pmosdelay}, we plotted the switch times with different gate-to-source-voltages for the NMOS and PMOS transistors.

From the figures, we can see that the best switch times appear at $5V$. More specifically, here we have a delay of approximately $30ns$ for both the turn-on and turn-off delay for the NMOS and a turn-on delay of approximately $10ns$ and approximately $25ns$ for the turn-off delay for the PMOS. 

Nevertheless, since we are also considering the power consumption, we will choose $V_{GS} = 3.3V$, where we have a turn-on delay of approximately $10ns$ and approximately $55ns$ for the turn-off delay for the NMOS. Furthermore, we have a turn-on delay of approximately $15ns$ and approximately $20ns$ for the turn-off delay for the PMOS.

Now we can calculate the latencies of the designed circuit. 
When both inputs of the NAND gate are high, we have a total delay time, which consists of the turn-on and turn-off delays,  of $t_d = 80ns $. If one is low and one high, we have a total delay time of $t_d = 80ns$; when both are low, we have a total delay time of $t_d = 80ns$. The highest delay is $80ns$ for one NAND gate. The delay then has to be multiplied by the depth of the Hadamard receiver, which is given by $\log2^n$, with $n$ being the order of the Hadamard receiver. For communication links of a length of $10$s or $100$s of kilometers, the propagation time of messages in the fiber is below or in the order of microseconds. In these scenarios, latency advantages are obtained from optical processing matter.

In comparison, the optical version of the FWHT has a little delay, as beamsplitters are passive devices, meaning that photons pass through at the speed of light without delay. The 20-mode photonic chip by QuiX Quantum \cite{Quix} has a size of $2 \times 3 cm$ with one tunable beamsplitter with the size of about $2mm$. As a result, to pass through a beamsplitter, it takes approximately $10ps$ and approximately $100ps$ through the entire chip. 

As mentioned before, we have to multiply the latencies of the beamsplitter and its electronic counterpart with the depth of the Hadamard transformation, which is given by $\log 2^n$. 

\section{Discussion}
This section will discuss the results obtained from our analysis of electrical and optical computation of the joint detection receiver, more specifically for the FWHT, and explore their implications for our research questions and objectives.
\subsection{Power Consumption}
Our analysis considered the power consumption of MOSFETs and beamsplitters. The MOSFETs, specifically the NMOS (SiRA04DP) and PMOS (SiA469DJ), exhibited different power consumption characteristics across their operating modes: cutoff, triode, and saturation. In cutoff mode, the MOSFETs consumed no power as they were not operational. The power consumption depended on the drain current and drain-source voltage in triode and saturation modes. We observed that the power consumption increased as the drain-source voltage increased, with the NMOS consuming higher power than the PMOS. However, at a gate-source voltage of $3.3V$, the power consumption was significantly reduced for both NMOS and PMOS compared to $5V$.
In contrast, beamsplitters, including the ones used in the Quixquantum photonic chip \cite{Quix}, were passive devices that did not consume power. Although the initial tuning of the beamsplitters required power, once tuned, they operated passively without consuming additional power.
\subsection{Decoding Latency}
The decoding latency of the chosen MOSFETs was determined by analyzing the turn-on and turn-off delay times. By adding a small external gate resistor, we achieved faster switch times. The results demonstrated that the best switch times were obtained at a gate-source voltage of $5V$. However, considering both power consumption and switch times, we selected a gate-source voltage of $3.3V$ for further analysis. At this voltage, the NMOS exhibited a turn-on delay of approximately $10ns$ and a turn-off delay of approximately $55ns$. At the same time, the PMOS showed a turn-on delay of approximately $15ns$ and a turn-off delay of approximately $20ns$.
In comparison, the optical version of the FWHT demonstrated minimal delay due to the passive nature of the beamsplitters. Photons passed through the beamsplitters at light speed without significant delay. For instance, the QuiX Quantum 20-mode photonic chip had a size of $2x3 cm$, with a single tunable beamsplitter measuring approximately $2mm$. The time required to pass through a beamsplitter was approximately $10ps$, and the total time to traverse the entire chip was approximately $100ps$.
Optical computation holds significant promise in terms of the latency advantages of optical processing, particularly for communication links spanning tens or hundreds of kilometers, where fiber propagation times are in the order of microseconds.

\section{Conclusions and Future Work}
In conclusion, our investigation into applying electrical and optical computation for the Fast Walsh-Hadamard Transform (FWHT), focusing on IoT scenarios, has provided valuable insights into power consumption and latency considerations. The power consumption analysis revealed a crucial trade-off between voltage levels and power usage, emphasizing the significance of lower gate-source voltages in reducing energy consumption. These findings guide selecting appropriate MOSFETs and operational voltages in IoT systems with specific power constraints.

Furthermore, the examination of decoding latency has yielded important information regarding switch times and their correlation with gate-source voltages. These results directly affect circuit design choices, enabling the optimization of electrical computation systems to meet the time-sensitive demands of IoT applications.

In contrast, optical computation, leveraging passive devices like beamsplitters, exhibited notably low latency, making it a compelling option for scenarios that require rapid processing and communication over extended distances. For future IoT research, exploring the scalability and integration of optical computation techniques becomes paramount, considering the complexities associated with the design and fabrication of photonic devices.

In light of these findings, our analysis of electrical and optical computation for the FWHT, when applied as a joint detection receiver in IoT applications, holds great promise for enhancing efficiency and performance. This research opens up exciting possibilities for implementing energy-efficient and time-critical IoT systems, making significant strides toward realizing a more connected and intelligent world.

\section{Acknowledgments}
Funding from the Federal Ministry of Education and Research of Germany, programme "Souver\ "an. Digital. Vernetzt." joint project 6G-life, project identification number: 16KISK002 (ZA, JN) and grant 16KISQ093 (JN), DFG Emmy-Noether program under grant number NO 1129/2-1 (JN) and support of the Munich Center for Quantum Science and Technology (MCQST) are acknowledged.
The project/research is part of the Munich Quantum Valley, supported by the Bavarian state government with funds from the Hightech Agenda Bayern Plus.

\end{document}

%% file: logic_bs.tex
\begin{tikzpicture}
% Input nodes
\node   (A1) at (0,2) {$A_1$};
\node   (A2) at (0,1) {$A_2$};
\node   (B1) at (0,-1) {$B_1$};
\node   (B2) at (0,-2) {$B_2$};
% AND gates
\node[draw, and gate US] (and1) at (2,1) {};
\node[draw, and gate US] (and2) at (2,0) {};
\node[draw, and gate US] (and3) at (2,-1) {};
\node[draw, and gate US] (and4) at (2,-2) {};
% Output nodes
\node   (C1) at (4,1) {$C_1$};
\node   (C2) at (4,0) {$C_2$};
\node   (D1) at (4,-1) {$D_1$};
\node   (D2) at (4,-2) {$D_2$};
% Connections
\draw (A1) -- (and1.input 1);
\draw (B1) -- (and1.input 2);
\draw (A2) -- (and2.input 1);
\draw (B2) -- (and2.input 2);
\draw (A1) -- (and3.input 2);
\draw (B2) -- (and3.input 1);
\draw (A2) -- (and4.input 2);
\draw (B1) -- (and4.input 1);
\draw (and1.output) -- (C1);
\draw (and2.output) -- (C2);
\draw (and3.output) -- (D1);
\draw (and4.output) -- (D2);
\end{tikzpicture}

%% file: pmosnmos.tex
\tikzstyle{orange}=[rectangle,fill=orange!10,inner sep=0mm,text=black,thick,draw=black!90]
\tikzstyle{blue}=[rectangle,fill=blue!10,inner sep=0mm,text=black,thick,draw=black!90]
\begin{circuitikz}[american,]

\draw (0,0)  node[pmos](PMOS){};
\draw (3,0) node[nmos](NMOS){};
\draw(-0.6,0.3) node[left](pgate){$G$};
\draw(0,0.5) node[right](psource){$S$};
\draw(0,-0.5) node[right](pdrain){$D$};

\draw(2.4,0.3) node[left](ngate){$G$};
\draw(3,0.5) node[right](ndrain){$D$};
\draw(3,-0.5) node[right](nsource){$S$};

\end{circuitikz}

%% file: and-gate.tex
\tikzstyle{orange}=[rectangle,fill=orange!10,inner sep=0mm,text=black,thick]
\tikzstyle{blue}=[rectangle,fill=blue!10,inner sep=0mm,text=black,thick]
\begin{circuitikz}[american,]

\node[orange,minimum width=5.8cm,minimum height=7.3cm] (bg) at (1,-1.5) {};
\node[blue,minimum width=3.5cm,minimum height=7.3cm] (bg) at (5.6,-1.5) {};

\draw (-2,-1) node[left](A){$V_A$};
\draw (-2,-1.5) node[left](B){$V_B$};

\draw (0,0)  node[pmos](PMOS1){PMOS1}  ;
\draw (2.5,0) node[pmos](PMOS2){PMOS2};

\draw (1,-2) node[nmos](NMOS1){NMOS1};
\draw (1,-3.5) node[nmos](NMOS2){NMOS2};

\draw (1,1) node[anchor=south](VDD1){$V_{dd}$};
\draw (5.5,1) node[anchor = south](VDD2){$V_{dd}$};

\draw (0.8,1) -- (1.2,1);
\draw (5.3,1) -- (5.7,1);
 
\draw (1,-4) node[ground](GND1){};
\draw (5.5,-4) node[ground](GND2){};

\draw (PMOS1.S) -- (PMOS2.S);
\draw (PMOS1.D) -- (PMOS2.D);
%\draw (NMOS1.S) -- (NMOS2.D);
\draw (NMOS1.D) -- (1,-0.8);
\draw (A) -| (PMOS1.B);
\draw (B)  -|  (PMOS2.B);
\draw (A) -| (NMOS1.B);
\draw (-1.5,-1.5)  |-  (NMOS2.B);
\draw (3.5,-1)  node(NANDOUT){};
\draw (2.8,-1) node[below](text){$V_{out_{NAND}}$};
\draw (1,-1) -- (NANDOUT);
\draw (5.5,0) node[pmos](PMOS3){PMOS3};
\draw (5.5,-2) node[nmos](NMOS3){NMOS3};
\draw (NANDOUT) -| (PMOS3.B);
\draw (NANDOUT) -| (NMOS3.B);
\draw (3,-1) -- (4,-1);
\draw (NMOS3.D) -- (PMOS3.D);
\draw(6.5,-1) node[below](ANDOUT){$V_{out_{AND}}$};
\draw(0,-5.5) node[right,text=orange](NANDgate){NAND gate};
\draw(4,-5.5) node[right, text=blue](NOTgate){NOT gate};
\draw (5.5,-1) -- (6.5,-1);
\draw(VDD1) -- (1,0.8);
\draw(VDD2) -- (5.5,0.8);
\draw (GND2) -- (NMOS3.S);

%\draw (Q1.B)|- (Q3.B);
%\draw (Q2.B) |- (Q4.B);
%\draw (-1,-1) -- (3,-1);
%\draw (Q1.D) to[short,-*] (0,-1);
%\draw (Q2.D) to[short,-*] (2,-1);
%\draw (Q4.D) to[short,-*] (2, -1);
%\draw (4.5,-1)  node[left](out){Output};

%\draw (Q3.S) node[pmos] (Q4) {q4};

%\draw (0,-2)  node[nmos](Q5){q5}  ;
%\draw (Q5.S)  node[nmos](Q6) {q6};
%\draw (2,-2) node[nmos](Q7){q7};
%\draw (Q7.S) node[nmos] (Q8) {q8};
%\draw (Q1.D) --(Q5.D);
%\draw (Q3.D) --(Q7.D);
%\draw (0,-1) to[short,*-*] (2,-1); 
%\draw (2,-1) -- (3,-1); 
%\draw (4.5,-1)  node[left](out){Output};

\end{circuitikz}